\documentclass[12pt,preprint]{aastex}
\usepackage{emulateapj5}
\usepackage{graphicx} 
\usepackage{latexsym}
\usepackage{dcolumn}
\usepackage{bm}
\input{epsf}
\newcommand{\bfig}{\noindent\begin{minipage}{3.48in}}
\newcommand{\efig}{\bigskip\end{minipage}}
\begin{document}
\title{Cross correlations of the cosmic infrared background}
\author{Pengjie Zhang}
\email{zhangpj@fnal.gov}
\affil{NASA/Fermilab Astrophysics Group,
Fermi National Accelerator Laboratory,
Box 500,
Batavia, IL 60510}
\begin{abstract}
The cosmic infrared background (CIB) is a sensitive measure of  the star formation history. But this background is
overwhelmed by foregrounds, which bias the CIB mean flux and auto
correlation measurement severely. Since dominant foregrounds do not
correlate with the large scale structure,  the cross correlation of CIB
with galaxies is free of  such foregrounds and presents as an unbiased
measure of the star formation history. In such cross correlation
measurement, one can utilize all frequency information, instead of
being limited to several narrow frequency windows.   This allows the
measurement of  CIB based on integrated intensity, 
whose theoretical prediction is  based on energy conservation, thus is fairly
model independent and robust.  The redshift information of CIB sources
can be directly recovered with the aid of  galaxy photometric
redshifts.  In the far IR region, correlation signal is around 
$10\%$ at degree scale. Combining FIRAS and SDSS, the cross
correlation can be measured with $20\%$ accuracy (statistical and
systematics errors) at $l\sim 40$. Such measurement would allow  the 
measurement of star formation rate with comparable accuracies to $z\sim
1$. Future CIB experiments with $1^{\circ}$ resolution are able
to measure the star formation rate with $10\%$ accuracy to $z\sim
1.5$. Secondary CMB anisotropies arising from the large scale structure
correlate with CIB. The correlation strength is about $0.5\mu$K at
$\sim 10^{\circ}$ (the integrated Sachs-Wolfe  effect) and $\sim
-0.3\mu$K at $\sim 
1^{\circ}$ (the Sunyaev Zeldovich effect, Rayleigh-Jeans regime).
FIRAS+WMAP would allow 
the measurement of the ISW effect with $\ga 5\sigma$ confidence. Its
detection  helps to constrain the amount of dark energy and far IR
emissivity bias. A future CIB experiment with better angular
resolution would allow the measurement of CIB-SZ cross correlation and
help to understand the effect of SN feedback on the SZ effect.
\end{abstract}
\keywords{cosmology: large scale structure; infrared: theory--diffuse
background; star: formation} 
\section{Introduction}
Detections of the cosmic infrared background (CIB) (refer to
\citet{Hauser01} for a 
recent review) open a new window for the study of star formation history. The main CIB component associates with star formation. Dust   
in  extragalactic galaxies  absorbs UV photons and reemits in far 
infrared region. Cosmic expansion shifts light emitted by low mass
population I/II 
stars or massive population III stars into near
infrared region. We denote this CIB originated from star formation as
SCIB, which 
provides a fairly unbiased and statistically robust measure of the star
formation history (SFH). But the precision CIB measurement is challenged by  
overwhelming foregrounds. Interplanetary dust emission dominates over
the CIB by an
order of magnitude at $\lambda\la 100 \mu$m.  Galactic interstellar
dust emission peaks at $\lambda \sim 100 \mu$m with 
an intensity at least comparable to the CIB. Bright galactic sources
such as stars and faint galactic sources have 
intensities at least several times larger than the CIB in near IR
region.  The CMB dominates over the $\lambda\ga 400\mu$m far IR region.

To reliably separate the CIB and  these dominant foregrounds, robust
understanding of foreground spacial and spectral distribution is required. The
removal of these foregrounds has enabled successful CIB detections. But the
residual foregrounds still cause large dispersions in the current CIB
intensity measurements (see table 1 of
\citet{Hauser01} for a review up to 
2001 and  \citet{Wright03}  
and reference therein for a recent compilation of COBE data). 

One way to bypass foregrounds is to investigate the CIB fluctuations, as
proposed by \citet{Bond86,Haiman00,Knox01}.  This fluctuation approach has
several 
advantages. (1) The CIB roughly follows star forming galaxy (SFG) distribution,
so one expects $\sim 10\%$ fluctuations at degree scale, which are about 4
orders   
of magnitude larger than the CMB fluctuations. (2) Fluctuations of galactic
foregrounds concentrate on large 
angular scales, for example, the power spectrum of the galactic interstellar
medium emission scales as $C_l\propto l^{-3}$ \citep{Wright98}. Thus
at sufficiently  small angular 
scales, The CIB fluctuations exceed these foreground fluctuations. 

But correlations of CIB foregrounds bias this approach. Fluctuations
of Galactic interstellar   dust emission 
dominate over the CIB intrinsic  fluctuations in a large sky fraction and 
large frequency and angular  ranges \citep{Knox01}.  The CIB auto
correlation measurement  in such regions requires an accurate removal of these
foreground correlations. The residual foreground correlations   bias the
intrinsic CIB auto correlation measurement and the following extraction of SFH.
This problem is most severe at large angular
scale ($l\lesssim 200$)
or short wavelength ($\lambda\lesssim 300 \mu$m) and may affect the estimation
of \citet{Knox01} significantly. To minimize such 
bias, the auto correlation measurement has 
to be limited to finite sky regions, frequency bands and angular
scales.  These limitations increase the sample variance.

For the purpose of extracting star formation rate (SFR),  various
non-stellar CIB, such as AGN CIB\footnote{We always refer AGN CIB as the CIB
contributed by AGN accretion.}, introduce extra systematic errors.
AGNs have similar clustering property as galaxies and similar 
thermal emission features as extragalactic dust, so it is hard to
separate AGN CIB and SCIB by  the 
mean flux and auto correlation measurements, even utilizing multi-frequency
information. Since 
AGN may contribute $\sim 10\%$ to the CIB intensities
(\citet{Hauser01} and reference 
therein), it may introduce $\sim 10\%$ systematic error to the SFR
measurement.

The ultimate challenge of the auto correlation  approach is to recover the
redshift information of CIB sources. This can be done  by 
multi-band CIB correlations \citep{Knox01}. But such approach requires
detailed understanding of dust composition and size distribution and
UV sources distribution, such modeling is difficult and 
uncertainty it introduces is hard to quantify.

With all these concerns, new CIB analysis method is
demanded. Dominant CIB foregrounds such as galactic foregrounds and 
primary CMB do not correlate with extragalactic galaxies.  This
suggests that crossing correlating the CIB with galaxies may alleviate
foreground contaminations. Residual
correlations originated from other CIB and extragalactic
contaminations are negligible comparing to the SCIB-galaxy
correlation.  Though AGN
CIB-galaxy cross correlation still exists in such approach and could
bias the result,  its effect is at most $\sim 10\%$. Furthermore,  one can
utilize the  difference of AGN redshift distribution versus that of SFGs to
at least partially separate the AGN contribution.
Thus, the cross correlation measurement of the CIB and galaxies is fairly
unbiased.  With the aid of
galaxy photometric redshift information, one does not  need to rely on
frequency information to recover redshift information of CIB
sources. We  only need to  model the 
integrated  CIB intensity, whose prediction is directly  based on  energy
conservation and thus fairly robust.   So the cross correlation of CIB
with galaxies would avoid most problems in the mean flux and auto
correlation approaches.

Since dust emission in far IR region is directly related to SFR, throughout
this paper, we focus on the far IR region. Similar method can be applied to
near IR region, which will constrain SFR at low mass end.  In
\S\ref{sec:SFIRB}, we estimate the auto and cross 
correlation power spectra of SCIB based on the integrated CIB intensity.  Our model
can  
be applied to CIB-lensing cross correlation, as \citet{Song03} did. But weak
lensing surveys are limited by finite sky coverage while redshift information
of 
lenses is entangled, so we postpone such calculation. We investigate the
observational 
feasibility of such cross correlation by estimating  the systematic
errors (\S\ref{sec:systematic}) and the statistical
errors (\S\ref{sec:error}).  Since the SCIB traces the large scale
structure, it correlates with secondary CMB anisotropies such as the integrated
Sachs-Wolfe (ISW) effect and the thermal Sunyaev-Zeldovich (SZ) effect. The
cross 
correlation of the CIB and the CMB provides another way to constrain
dark energy and the 
large scale structure. We predict the strength of such correlation in \S
\ref{sec:CMB-CIB}. Our predictions are based on the  WMAP-alone cosmology with
$\Omega_m=0.268$, $\Omega_{\Lambda}=1-\Omega_m$,
$\Omega_b=0.044$, $\sigma_8=0.84$ and 
$h=0.71$ \citep{Spergel03}\footnote{For simplicity, we
adopt a slightly different $\Omega_{\Lambda}$ from the best fit value
$1.02-\Omega_m$. This change has negligible effect on our results.}.

\section{Intrinsic cosmic infrared background fluctuations}
\label{sec:SFIRB}
We model the SCIB based on energy conservation.  Observations suggest that
$f_{\rm absption} \sim 90\%$ UV light is 
absorbed by dust 
(\citet{Massarotti01a} and reference therein) and reemitted in far IR.  Since
UV 
 emission is dominated by short-life O/B stars, UV
emissivity $\rho_{UV}$ is directly related to SFR by
\begin{equation}
\rho_{\rm UV}(\nu)=\alpha_{\nu} \frac{\rm SFR}{M_{\sun}{\rm yr}^{-1}}.
\end{equation}
Coefficient $\alpha_{\nu}$ is determined by the initial mass function
(IMF). It strongly depends on the shape of IMF at high mass end. At $1500$ \AA,
for the Salpeter IMF, $\alpha_{\nu}\simeq 8.0\times 10^{-25} \ {\rm nW}\  {\rm
Hz}^{-1}$\citep{Madau98}. For the Scalo IMF which has much less
massive stars, $\alpha_{\nu}$ is about twice smaller. Since the  Scalo IMF
produces too red  integrated galaxy spectra \citep{Lilly96} and too
low CIB intensity,  we consider only the Salpeter IMF in this paper. For
the  Salpeter  IMF, $\alpha_{\nu}$  has only a weak dependence on $\nu$.  For
example, at
$2800$ \AA, $\alpha_{\nu}\simeq 7.9\times 10^{-25}\ {\rm
nW}\ {\rm Hz}^{-1}$. Then the
integrated dust FIR emissivity is given by 
\begin{eqnarray}
\bar{j}_{d}&=&\alpha f_{\rm absption}  \frac{\rm SFR}{M_{\sun}{\rm
yr}^{-1}}.
\end{eqnarray}
Here, $\alpha=\int_{\rm UV}\alpha_{\nu}d{\nu}\simeq 1.6\times 10^{-9} {\rm
nW}$.  The integrated SCIB  intensity at direction $\hat{n}$ is 
\begin{equation}
I(\hat{n})=\int \frac{\bar{j}_{d}(1+\delta_j(\chi,\hat{n}))}{4\pi(1+z)^2} d\chi.
\end{equation}

Here, $\delta_j$ is the fractional fluctuation of the integrated IR
emissivity. $\chi$ is the comoving distance. SFR at $z\lesssim 1.2$ is
relatively 
robustly measured and we adopt it 
as ${\rm SFR}(z)=10^{-2.1} \exp(t/2.6{\rm Gyr}) M_{\sun} {\rm yr}^{-1} 
{\rm Mpc}^{-3}$ \citep{Hippelein03}.   Here, $t$ is the look-back time. 
For $z>1.2$, we consider three SFR:
\begin{eqnarray}
{\rm SFR}(z=1.2) \exp(-\frac{t-t(z=1.2)}{2.5{\rm Gyr}})&:\  {\rm SFR}1
\nonumber \\ 
{\rm SFR}(z=1.2)&:\ {\rm SFR}2 \\ \nonumber
10^{-2.1}\exp(\frac{t}{2.6{\rm Gyr}}) M_{\sun} {\rm yr}^{-1} 
{\rm Mpc}^{-3} &:\  {\rm SFR}3 \ .
\end{eqnarray}
For the Salpeter IMF, we obtain $\bar{I}\simeq 16, 20, 30 \ {\rm nW}\  {\rm m}^{-2}\ {\rm
sr}^{-1}$, respectively. These results are consistent with observations,
but due 
to the large dispersion in the observational results (see
\citet{Hauser01,Wright03} for reviews),
no specific conclusion can be drawn. 

We  calculate the CIB fluctuations by the Limber's equation: 
\begin{equation}
\frac{l^2}{2\pi}C_{\rm IR}\bar{I}^2=\pi \int
 \left(\frac{\bar{j}_d}{4\pi(1+z)^2}\right)^2 \frac{\chi}{l}\Delta^2_j(\frac{l}{\chi},z)
 d\chi,
\end{equation}
\begin{equation}
\label{eqn:CjG}
\frac{l^2}{2\pi}C_{\rm IR,G}\bar{I}\bar{\Sigma}_G=\int_{z_1}^{z_2}
\frac{\bar{j}_d}{4(1+z)^2} \frac{\chi}{l}\Delta^2_{jG}(\frac{l}{\chi},z)
 \frac{dn}{dz} dz.
\end{equation}
Here, $\Delta^2_j$ and $\Delta^2_{jG}$ are the variances (power spectra) of
$\delta_j$ and its cross correlation with galaxy number overdensity $\delta_G$, respectively.
$dn/dz(z)$ is the galaxy number distribution
function in a given galaxy survey. The galaxy surface density is defined as
$\Sigma_G=\int dn/dz 
(1+\delta_G) dz$. We adopt the parametric form 
$dn/dz=3z^2/2/(z_m/1.412)^3\exp(-(1.412z/z_m)^{1.5})$ \citep{Baugh93}. $z_m$
is the median redshift of the galaxy distribution. For SDSS, We
$z_m\sim 0.5$ \citep{Dodelson02}. The above equations hold at
angular scales $\lesssim 30^{\circ}$ or $l\ga 10$. For smaller $l$, one has to
calculate the spherical Bessel integral. Since the small $l$ modes are
dominated by 
cosmic variance and/or foregrounds, whose power concentrates on large angular
scales, we neglect the calculation for such small $l$.   
\bfig
\plotone{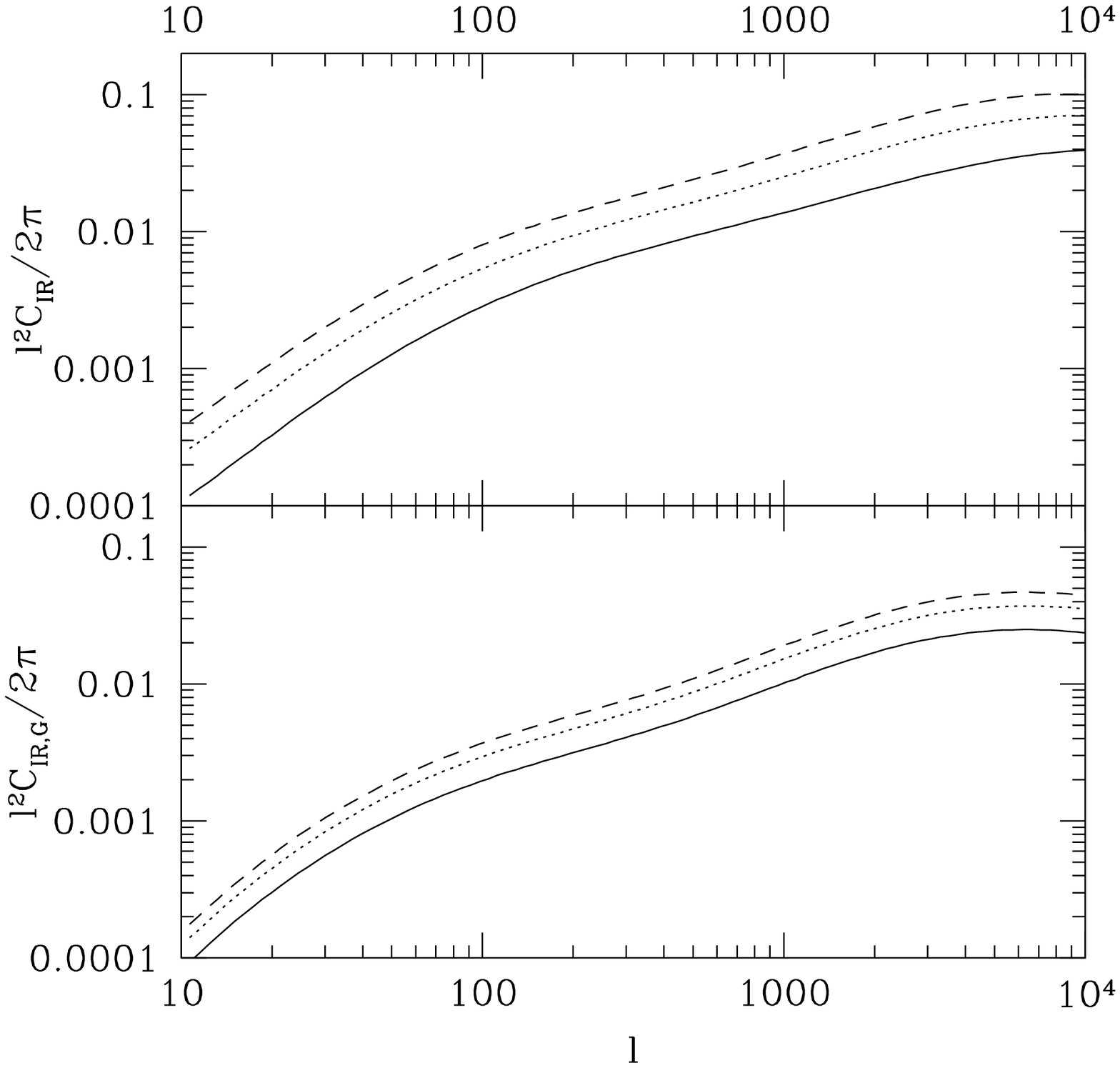}
\figcaption{The CIB power spectra. The top panel plots the CIB auto correlation
power spectra while the bottom panel plots the cross correlation power spectra
with SDSS 
galaxies. We assume three SFR: SFR1 (dash lines), SFR2 (dot lines) and SFR3
(solid lines). The change of amplitude reflects the change of redshift where
the dominant correlation signal comes from. }
\efig
$\Delta^2_j$ and $\Delta^2_{jG}$ are generally unknown. Since dust
associates with SFGs, averaging over many galaxies, $j_d$ should trace SFG
distribution, which in turn traces dark matter distribution.  A $10^{'2}$ sky
area contains more than $100$ SFGs, so one expects that the constant
bias assumption $\delta_j=b_j \delta_{\rm DM}$ applies  at $\ga 3^{'}$
scale. In principle, $b_j$ is an observable and can be inferred from
the combined CIB mean flux and correlation measurements (\S
\ref{sec:error}) or from 
the CIB-ISW cross correlation measurement (\S \ref{sec:CMB-CIB}).  For
simplicity, we assume  $b_j=b_{\rm SFG}=3$ \citep{Haiman00}. This SFG
bias is inferred from Lyman break galaxies \citep{Giavalisco98}. The dark
matter power spectrum is calculated by the code of \citet{Smith03}.

The fractional fluctuations at degree scales are about $10\%$ at degree
scale, which 
translate to intensity fluctuations $\sim 2\  {\rm nW}\  {\rm m}^{-2}\  {\rm
sr}^{-1}$.  These fluctuations are consistent with both theoretical predictions
\citep{Haiman00,Knox01} and observations \citep{Hauser01,Miville-Deschenes02}. 

\section{Error estimation}
The above prediction assumes that the observational wavelength band contains
all extragalactic dust emission. The requirement on the wavelength band can be
estimated as follows.  
Extragalactic dust  emission can be described  by a 
gray-body spectrum $I_{\nu}\propto \nu^{\beta}B_{\nu}(T)$ with  typical
$T\sim 20$ K (see, e.g., \citet{Haiman00}) and $\beta=2$ \citep{Draine84}.
Here, $B_{\nu}(T)$ is the 
Planck function.  $\nu I_{\nu}$ peaks at $\simeq 2.5 T/20 {\rm K}$ THz (or,
equivalently, $20 {\rm K}/T 120 \mu $m).  More than $96\%$ emission comes from
the range $[0.8,5] T/20 
{\rm K}$ ($[60,375]20 
{\rm K}/T \mu$m).  Since star formation rate was more than an order of
magnitude higher at $z\sim 1$ than at present, considering the redshift
effect, an integral range $[0.3,3]$ 
THz ($[100,1000]\mu$m) should contain almost all contribution from
extragalactic dust emission. 
Indeed, from  the fitted FIRAS far infrared background (FIRB) spectrum \citep{Fixsen98} 
\begin{equation}
I_{\nu}=1.3\pm0.4\times 10^{-5}(\frac{\nu}{3{\rm THz}})^{0.64\pm0.12}
B_{\nu}(18.5\pm 1.2 {\rm K}),
\end{equation}
this fraction is $\ga 92\%$. This frequncy range may underestimate low
redshift contribution by $10\%$, but it contains $\ga 95\%$ extragalactic dust
emission from $z\ga 0.6$ and is  sufficient for our extraction of SFR at $z\sim
1$.  

FIRAS satisfies this wavelength band requirement, so we take it as our target
to estimate both the statistical  and systematic errors of the cross
correlation measurement.  The assumption  that $[0.3,3]$ THz contains all dust
emission  would introduce several percent errors at low redshift, which are minor comparing to other
systematic errors (\S 
\ref{sec:systematic}).

\subsection{Systematics errors}
\label{sec:systematic}
Non-stellar CIB sources, if correlate with large scale structure, bias SFR
measured from the CIB-galaxy cross correlation. In this section, we
estimate the systematic errors
caused by such sources. We find that the dominant systematic error comes from
AGN CIB, which may contribute $10\%$ to the cross correlation. We show that
AGN CIB can  be at least partially subtracted. In the worst case that
these sources are not 
subtractive, they can introduce  at most $10\%$ systematic error to the measured $\bar{j}_d$, which is still
quite accurate comparing to current SFR measurement (see,
e.g. \citet{Hippelein03} for recent data compilation).  
\subsubsection{AGN}
AGN may contribute $\sim 10\%$ to the CIB flux through the thermal
emission of dust 
heated by the central black hole  (\citet{Hauser01} and reference
therein). This non-stellar AGN 
contribution would affect the determination of SFR by a comparable
fraction. AGNs have similar clustering property as galaxies and similar
thermal emission feature as extragalactic dust, so it is hard to
separate AGN CIB and SCIB by the 
mean flux and auto correlation measurements, even with the aid of
multi-frequency information. Many high redshift AGNs present as luminous
infrared galaxies  and dominate 
$L_{\rm IR}>10^{11} L_{\sun}$ object catalog \citep{Sanders96}.  So point
source identification, at least at high luminosity end, may be feasible and
help to understand the contribution of AGNs.  But the distinguishing of 
energy sources of such IR emission, gravitational potential energy or nuclear
energy, is not very clear (\citep{Hauser01} and reference therein).

AGN contamination can be partially cleaned by cross correlation with galaxies.
\begin{itemize}
\item Dust heated by starbursts has temperature $T\sim 20$K  while dust associated with AGN IR emission has temperature
$T\sim 60$-$100$ K \citep{Haas98}. Since the observed Far IR CIB
corresponds to 
gray-body  radiation with $T\sim 18$K,  the AGN CIB mainly comes from high
redshifts. Utilizing the redshift 
information obtained from  the cross correlation with galaxies, AGN
contamination could be  localized to high $z$.  
\item AGN-late type galaxy
cross correlation is much weaker than AGN-early type galaxy cross correlation
at arc minute scale \citep{Brown01} while SFGs strongly
correlate with late type galaxies. For future CIB survey which can resolve
this scale, one  can distinguish the contribution of AGN by cross
correlating CIB with early and later type galaxies separately.
\end{itemize}

\subsubsection{The integrated Sachs-Wolfe effect}
The time variation of gravitational potential field introduces secondary CMB
temperature fluctuations \citep{Sachs67}:
\begin{equation}
\Theta_{\rm ISW}\equiv\frac{\delta T}{T_{\rm CMB}}=2\int \frac{\dot{\phi}}{c^2}a\frac{d\chi}{c}.
\end{equation}
If the variation of gravitational potential $\phi$ is caused by the deviation
from a  $\Omega_m=1$ universe, this effect  is called the integrated
Sachs-Wolfe (ISW) effect.
The contribution of the ISW effect to CIB intensity is 
\begin{equation}
 I_{\rm ISW}=\Theta_{\rm ISW} \int_{\rm FIR} \frac{\partial I_{\nu}}{\partial
\nu}d\nu\simeq \Theta_{\rm ISW}0.2 \ {\rm nW}\  {\rm m}^{-2}\  {\rm sr}^{-1}.
\end{equation}
The temperature fluctuation caused by ISW $\delta \Theta_{\rm
ISW}$ is $\lesssim 10^{-5}$, so fluctuations of the CIB caused by ISW are $\sim
10^{-6} \ {\rm nW} \ {\rm m}^{-2} \ {\rm sr}^{-1}$. We then can safely neglect
this effect.

\subsubsection{The thermal Sunyaev-Zeldovich effect}
Free electrons scatter off CMB photons by their thermal motions and introduce
secondary CMB temperature fluctuations, namely, the thermal
Sunyaev-Zeldovich (SZ)  effect. It 
contributes an intensity $I_{\rm SZ}$ to  the CIB \citep{Zeldovich69}
\begin{eqnarray}
I_{\rm SZ}&=&-2 y \int_{\rm FIR}
\frac{x\exp(x)}{\exp(x)-1}\left(2-\frac{x/2}{\tanh(x/2)}\right)d\nu\\
\nonumber&\sim& y 10^3 \ {\rm nW}\  {\rm m}^{-2}\  {\rm
sr}^{-1}.
\end{eqnarray}
Here $x\equiv h\nu/k_BT_{\rm CMB}$. Since electron thermal energy is much
higher than CMB photon energy, photons gain energy and shift toward infrared
frequency after 
scattering. So the SZ contribution to the CIB is much larger than that
of the ISW  effect. 

The {\it y} parameter is given by the integral of electron thermal pressure
$p_e$ 
along the line of sight 
\begin{equation}
y=\int \frac{n_e k_B T_e}{m_ec^2} \sigma_T a d\chi\equiv\int \frac{p_e}{m_ec^2} \sigma_T a d\chi.
\end{equation}
The fluctuation of {\it y} is $\sim 10^{-6}$ at degree scale (see, e.g.,
\citet{Zhang01,Zhang02}).  Then fluctuations it introduces to the
CIB are  $\sim 
10^{-3} \ {\rm nW}\  {\rm m}^{-2}\  {\rm sr}^{-1}$, which are still $\sim 1000$
times 
smaller than SCIB fluctuations and can be neglected.

\subsubsection{Brown drawfs}
Brown drawfs associate  with galaxies, so the CIB caused by brown
drawfs has fractional fluctuations of the same order as the SCIB. Since the total
mass of brown drawfs accounts for $\ll 1\%$ of the total energy in the
universe, its contribution to CIB intensity  is $\ll 0.01 {\rm nW} {\rm m}^{-2}
{\rm 
sr}^{-1}$ \citep{Karimabadi84}. So its contribution to CIB fluctuations is
at most $\sim 0.1\%$ of SCIB fluctuations and can be neglected.   

\subsubsection{Intergalactic dust}
Intergalactic dust correlates with galaxies. The CIB contributed by
intergalactic dust should have similar fractional fluctuation as the
SCIB. But the 
contribution of intergalactic dust to the CIB flux seems to be
negligible based on SCUBA  measurement (see \citet{Hauser01}
and reference therein for detailed discussion). The existence of abundant
intergalactic dust would dim  faraway SNIas and alter the conclusion of
accelerating expansion of the universe based on SNIa Hubble diagram. Since WMAP
has independently  confirmed the acceleration of the  expansion of the
universe
without SNIa prior \citep{Spergel03}, there is little room left for
intergalactic dust.  So we will neglect this possible source. 

\subsubsection{Other cosmic sources}
\citet{Bond86} discussed several other possible CIB sources.  Primeval
galaxies, first stars and decaying particles all reside at very high
redshifts. Gravitational lensing caused by low z dark matter (shallower
than the corresponding galaxy survey) correlate the high z emission
with low z galaxies. But such correlation is 
negligible due to two reasons. Firstly, the correlation is caused by low
z dark matter lensing. For example, the cross correlation with SDSS would be
mainly caused by dark matter at $z\sim z_m/2=0.2$. So the lensing
effect is small. Secondly, these high z CIB sources only contribute a
small fraction, if not negligible,  to the total CIB intensity.

\subsubsection{Contaminations of galaxy surveys}
Galactic dust causes extinction, which biases the observed galaxy
distributions. This effect may correlate the  galactic dust foregrounds
of CIB with extragalactic galaxies. But since at least the first order
effect of extinction  is included in the observed galaxy number
distribution $dn/dz$, one  expects the residual correlation
caused by dust extinction to be negligible. Stellar contamination in
galaxy surveys also
correlate CIB foregrounds with galaxy distribution. But such
contamination is at $1\%$  level (e.g. SDSS, \citet{Sheldon03}) and is
thus negligible, too.

 \subsection{Statistical errors}
\label{sec:error}
The statistical errors of $C_{\rm IR,G}$ measurement are
\begin{equation}
\frac{\Delta (C \bar{I}\bar{\Sigma}_G)}{C
\bar{I}\bar{\Sigma}_G}=\sqrt{\frac{1+(C_{\rm IR}+C_{\rm
dust}+\frac{C_{\rm
short}}{W_l^2})(C_G+\frac{C_{N,G}}{W_l^2})C^{-2}_{\rm
IR,G}}{(2l+1)\Delta l f_{\rm sky}}}. 
\end{equation}
In this expression, we have assumed Gaussian fluctuations in the CIB,
which should at least hold at degree or larger scales. 
 In the $[0.3,3]$ THz frequency range, the integrated CMB intensity fluctuation
is
negligible. We consider two dominant noise 
sources: interplanetary and galactic dust, and short noise of the signal. We
adopt $C_{\rm dust} =  ( 10\ {\rm nW m^{-2} sr^{-1}})^2 l^{-3}$. The
$l$ dependence is adopted from \citet{Wright98} and the amplitude roughly
corresponds to the case avoid of the galactic plane
\citep{Wright98}. The short noise power spectrum is estimated by $C_{\rm
short}=4\pi/N_{\rm SFG}$. In this expression, we have omitted the flux
variation of individual SFG.  We further assume the total number of
SFG to be $N_{\rm
SFG}=2\times10^9$. The window function $W_l$ reflects the angular resolution of
the CIB survey. We approximate it as a Gaussian function
$W_l=\exp(-l^2\theta_0^2/2)$ with $\theta_0={\rm
FWHM}/(2\sqrt{2\ln{2}})$. While the FIRAS 
beam FWHM is nominally $7^{\circ}$ across it has sharper edges than a Gaussian beam
so for many applications it is better approximated by a Gaussian beam
with $5^{\circ}$ FWHM\footnote{Private communication with Dale J. Fixsen.}.
$C_{N,G}=4\pi f_{\rm sky}/N_G$ is the
Poisson noise of a galaxy survey. We adopt the 
expected number of SDSS galaxies as $N_G=5\times 10^7$ and choose $f_{\rm
sky}=1/4$. The bin size $\Delta l$ is chosen to be $\Delta l=0.5 l$.      
\bfig
\plotone{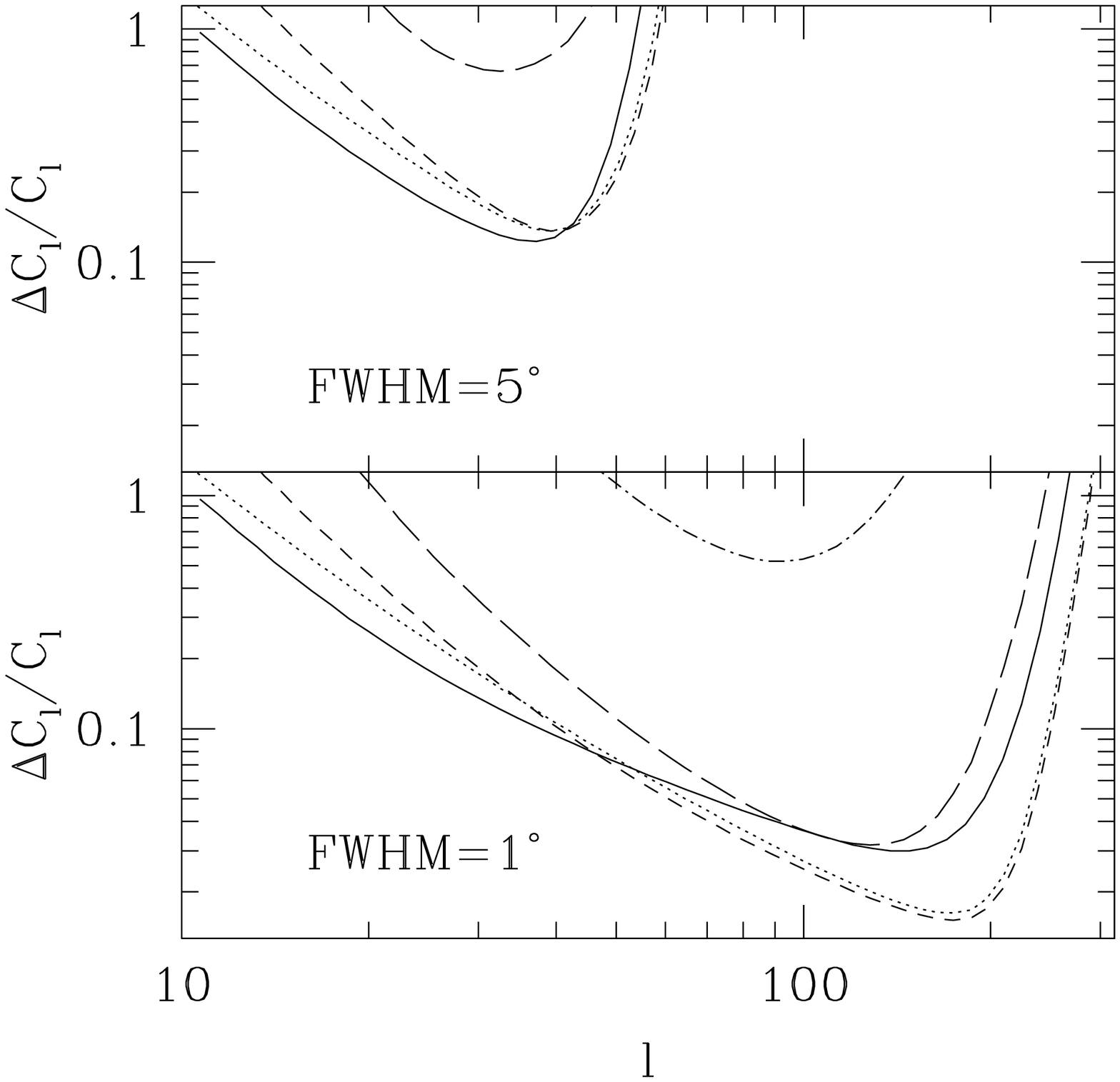}
\figcaption{The statistical errors in the CIB-galaxy correlation measurement. The redshift bins adopted are: $[0, \infty]$ (solid line),
$[0.0,0.2]$ (dot line), $[0.4,0.6]$ (short dash line), $[1.0,1.2]$ (long
dash line) and $[1.6,1.8]$ (dot-dash line). The top panel corresponds
to FIRAS+SDSS and the bottom panel corresponds to a toy CIB
experiment with $1^{\circ}$ FWHM. The cross correlation measurement is
mainly limited by the angular resolution of CIB experiment at low $z$
and limited amount of galaxies at high $z$. \label{fig:clerror}}
\efig
$C_{\rm IR,G}$ can be measured with $\sim 10\%$ accuracy at
$10^{\circ}$ scales and $z\la 0.6$ for a 
combination of FIRAS and SDSS (Fig. \ref{fig:clerror}). In these
ranges, the error of cosmic variance dominates. To beat down the
cosmic variance, higher angular resolution CIB experiment is
required, which can probe smaller angular scales
(Fig. \ref{fig:clerror}). Since the median 
redshift of  SDSS galaxies is $z_m\sim 0.5$, where most correlation signal
should come from, one expects the optimal measurement to be at $z\sim 0.5$
(Fig. \ref{fig:clerror}). Since only a small fraction of SDSS galaxies lies at
$\ga 1$, the correlation signal is weak and the short noise of galaxy
distribution is large. Thus the correlation measurement becomes noisy. To measure
the cross correlation at $z\ga 1.5$, a deeper galaxy survey is needed.
\bfig
\plotone{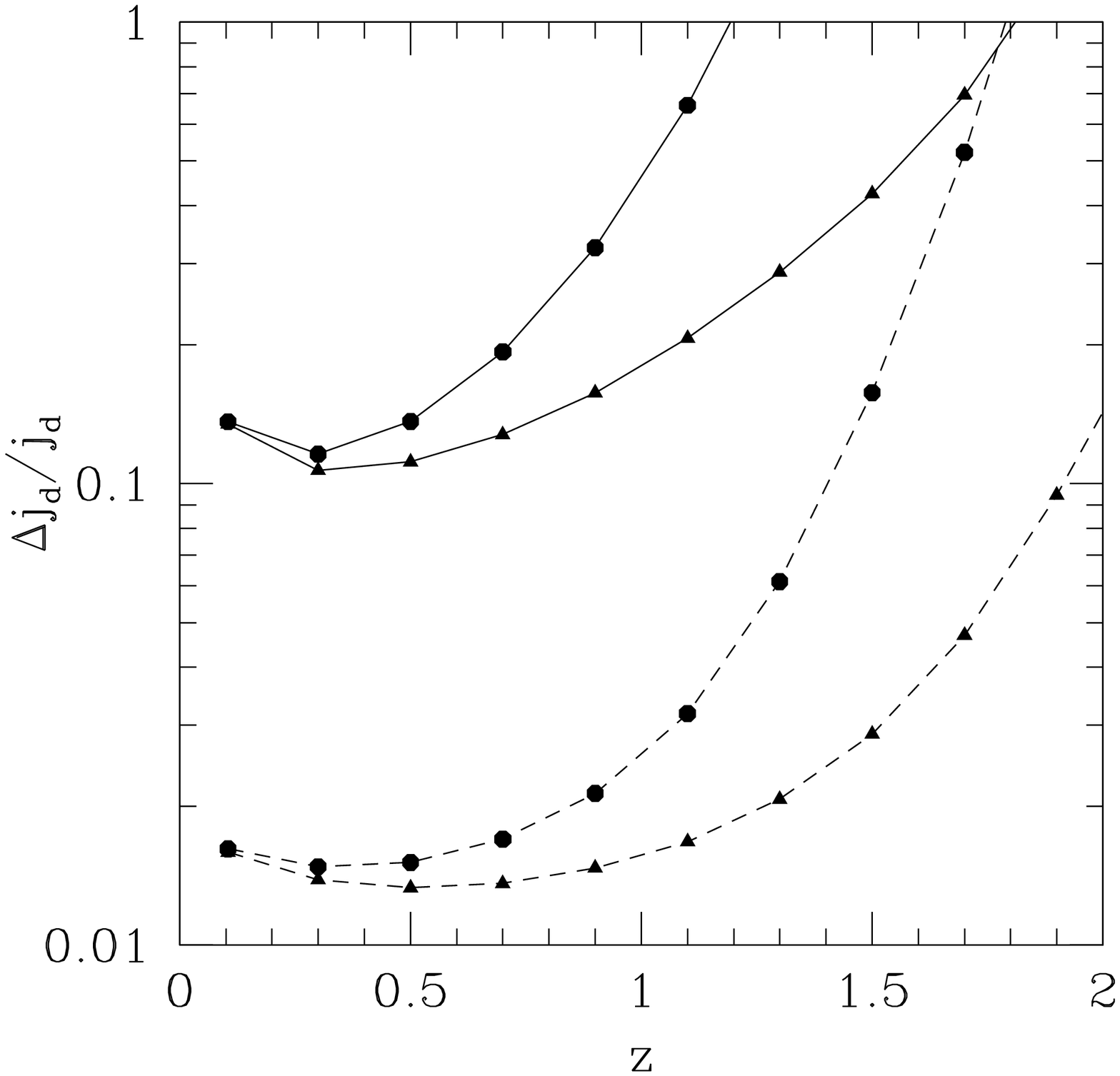}
\figcaption{The estimated statistical errors of $\bar{j}_d$ measured from
the CIB-galaxy  correlation. Data points connected with solid lines
correspond to FIRAS and those connected with dash lines correspond
to a toy CIB experiment with $1^{\circ}$ FWHM. Filled
circle points correspond to SDSS with a Galaxy median redshift
$z_m=0.5$ and filled 
triangle points corresponds to a toy galaxy survey with
$z_m=1.0$ and the same amount of galaxies at low redshift as SDSS. We
have assumed the  3D correlation of IR emissivity-galaxy to be known. \label{fig:jerror}}
\efig

Given $\Delta^2_{jG}$,  $\bar{j}_d(\bar{z})$ can be directly inferred
from $C_{\rm IR,G}$ (Eq. \ref{eqn:CjG}). The accuracy of the  measured
$\bar{j}_d(\bar{z})$ is then  
determined by the accuracy of $C_{\rm IR,G}$ measurement. 
 We approximate the relative error of the inferred $\bar{j}_d(z)$ as the
minimum $\Delta C_{\rm IR,G}/C_{\rm IR,G}$. The result estimated by this
somewhat arbitrary simplification should not deviate  from that of a
sophisticated analysis by a factor $\sim 2$ (Fig. \ref{fig:clerror}), so 
it is sufficient for our 
purpose. Currently, the extraction of $\bar{j}_d(z)$ is mainly limited
by the finite 
depth of the galaxy survey (Fig. \ref{fig:jerror}). FIRAS+SDSS could measure
the dust emissivity $\bar{j}_d$ with $\sim 10\%$ accuracy at $z\la 0.7$. For
higher $z$, the measurement becomes noisy quickly. The accuracy of $\bar{j}_d$
measurement is limited by the angular resolution of CIB experiment at
low $z$ and the 
amount of galaxies in galaxy surveys at high $z$. Combining a possible
future CIB experiment with $1^{\circ}$ FWHM and a deeper galaxy survey
with median redshift $z_m=1.0$, one may probe $\bar{j}_d$ and the star
formation history to $z\sim 1.5$ with several percent statistical errors. 

Since our redshift bins are $\Delta z\sim 0.2$ and one does not expect
strong evolution of $\Delta^2_{jG}$ over these redshift ranges, the
above analysis does not require precision measurement of galaxy redshifts. Photometric redshift can be measured with  $\sigma_z/(1+z_{\rm spec})\sim 0.04$
at $z<6$ \citep{Massarotti01b}, so  the error of photo-z should not affect our
result significantly.

The extraction of $\bar{j}_d$ is based on known $\Delta^2_{jG}$, which
is in principle an observable. Since at sufficiently large scale, the
cross correlation coefficient of $j_d$ and 
galaxies is unity, combining the mean CIB flux and CIB auto  correlation 
measurement and galaxy $\Delta^2_G$ measurement, $\bar{j}_d$ and
$\Delta^2_{jG}$ can be determined 
simultaneously given the cosmology. However, the 
feasibility of this approach relies on future accurate removal of foregrounds,
so we  postpone its discussion in this paper. Furthermore, as we will show in
\S \ref{sec:CMB-CIB}, the measurement of CIB-ISW cross correlation is
able to 
determine the mean bias of $j_d$ and thus $\Delta^2_{jG}$.

\section{CIB-CMB cross correlation}
Recent successful detections of the cross correlation between WMAP and galaxies
\citep{Fosalba03a,Fosalba03b,Scranton03,Afshordi03} present a powerful
way to constrain 
cosmology, especially dark energy (through the ISW effect). But current galaxy
surveys are  either limited by finite sky coverage (For APM, $  4300\
{\rm deg}^2$  and for present SDSS, $3400\  {\rm deg}^2$) or by
finite survey depth (For 2MASS, $\langle z\rangle\la 0.1$). This
limits current  detections to $\la 3\sigma$ level. To bypass these
limitations, one can  cross correlate CMB with other cosmic
backgrounds, to reduce cosmic variance and  
to amplify cross correlation signal. One choice  is the
X-ray background (XRB). But the measurement of WMAP and the all sky XRB survey
ROSAT fails to show any  significant correlation \citep{Diego03}. On one hand, XRB is dominated by AGNs,
which mainly concentrate to high redshift. On the other hand, ISW mainly comes
from low redshift where $\Omega_m$ deviates significantly from $1$. At degree
scale, which is the finniest scale WMAP+ROSAT can probe, SZ contribution also 
mainly comes from low redshift (see, e.g., \citet{Zhang01}). So one does not
expect a strong correlation between WMAP and ROSAT. Another choice is the
cosmic radio background (CRB). Since the Milky Way is a strong synchrotron
radiation source, which has a similar spectral index as the CRB synchrotron
component, the subtraction of CRB foregrounds is difficult. Without a
robust subtraction of CRB foregrounds, the tight correlation between
galactic foregrounds (For CMB, synchrotron and free-free emission,
\citet{Bennett03})  may bias the  
CIB-CMB correlation  measurement significantly. On the other hand, the origin of CRB is difficult to predict
analytically. The synchrotron component of CRB associates with magnetic
field, which is poorly  understood. The thermal component of CRB associates
with HII regions generated by massive star formation and is hard to predict
too. So, even if the intrinsic cross correlation is detected, it would be hard
to 
interpret. MeV cosmic $\gamma$-ray background (CGB) is expected to have a tight
correlation with the large scale structure, but current CGB
experiments may be  too noisy to detect any significant cross correlation with
CMB \citep{Zhang03a}.

SCIB traces star forming galaxies, which have strong 
correlation with gravitational potential (ISW) and gas thermal pressure (SZ
effect). The modeling of SCIB is
relatively straightforward, as discussed in previous sections. So, the
interpretation of CIB-CMB correlation is relatively robust. Thus,
SCIB-CMB cross correlation is a good way to measure the ISW effect and
the SZ effect. The detection of such cross correlation signal is
observationally feasible, as implied by the successful detection of
COBE-FIRAS correlation \citep{Burigana98}. 
\bfig
\plotone{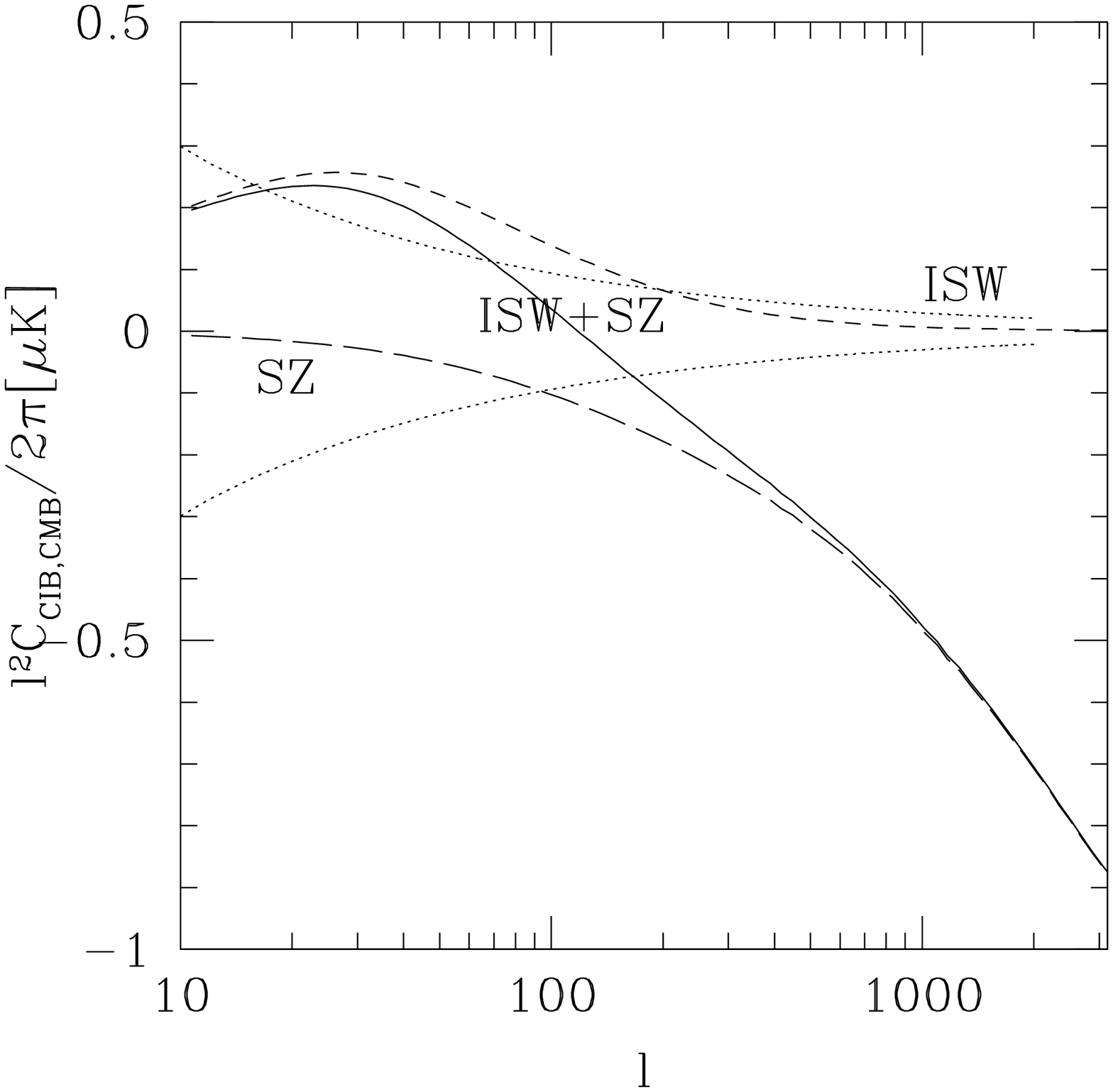}
\figcaption{CIB-CMB cross correlation. The SZ effect is estimated in the
Rayleigh-Jeans regime. The dot lines are the upper limit of systematic
errors caused by the CIB and the CMB dust foreground correlation. This
estimation 
assumes  no foreground reduction in CMB maps and assumes perfect foreground correlation.  With
reasonable CMB foreground removal, the systematic error can be reduced
by a factor of 10. \label{fig:CIB-CMB}}
\efig
\label{sec:CMB-CIB}
\subsection{CIB-ISW cross correlation}
The cross correlation power spectrum of ISW-CIB is given by 
\begin{equation}
\frac{l^2}{2\pi}C_{\rm IR,ISW}\bar{I}=\int
 \frac{\bar{j}_d}{4(1+z)^2} \frac{2 a}{c^3}
 \frac{\chi}{l}\Delta^2_{j\dot{\phi}} (\frac{l}{\chi},z)
 d\chi.
\end{equation}
If assuming a bias model ($\delta_j=b_j\delta$), we obtain
\begin{eqnarray}
\Delta^2_{j\dot{\phi}}(k,z)&=&\frac{3}{2}H_0^2b_j\left(\frac{d}{dt}(a^{-1}
\frac{\Delta^2_{\delta}}{k^2})-a^{-1}
\frac{\dot{\Delta^2_{\delta}}}{2k^2}\right) \\ \nonumber
&\simeq&\frac{3}{2}H_0^2 b_j\frac{d}{dt}(\frac{D}{a})
\frac{\Delta^2_{\delta}}{Dk^2} \ {:\ \rm linear\  regime}
\end{eqnarray}
Here, $D$ is the linear density growth factor, calculated by the fitting
formula of \citet{Carroll92}.   $a$ is the scale factor. $H_0$ is the present
Hubble constant. The power of $\Delta^2_{j\dot{\phi}}$ concentrates on large
scale due to the 
$k^2$ term in the denominator, thus the ISW effect dominates the CIB-CMB correlation
at large angular scale 
($l\lesssim 100$) and drops to zero quickly toward small scale (Fig. \ref{fig:CIB-CMB}). The
correlation amplitude reaches  $\sim 0.5 \mu$K at $l\sim 20$. One bonus of 
ISW-CIB correlation is that it gives the mean $b_j$ averaged over redshift.   
\subsection{CIB-SZ cross correlation}

The CIB-SZ cross correlation power spectrum in the Rayleigh-Jeans regime  is given by 
\begin{equation}
\frac{l^2}{2\pi}C_{\rm IR,SZ}\bar{I}=-2 \int
 \frac{\bar{j}_d}{4} \frac{\bar{p}_e}{m_ec^2}
 \sigma_T 
 \frac{\chi}{l}\Delta^2_{jp}(\frac{l}{\chi},z)
 d\chi.
\end{equation}
Here, $\Delta^2_{jp}$ is the cross correlation power spectrum (variance) of
$\delta_j$ and $\delta_p\equiv p_e/\bar{p}_e-1$.  We
calculate the
gas density weighted temperature $\bar{p}_e$ by the continuum field model
\citep{Zhang01}, which agrees with $\Lambda$CDM and self similar hydro
simulations very well \citep{Zhang03b}. This model predicts a mean temperature
decrement $5.0\mu$K. The modeling of $\Delta^2_{jp}$ is much more
complicated. For simplicity, we assume $\delta_p$  traces dark matter
overdensity with a constant bias $b_p=5$, as predicted by \citet{Zhang01}.     
The CIB-SZ correlation dominates at $l\ga 100$. At one degree
scale ($l\sim 300$),  the cross correlation amplitude reaches $-0.3\mu$K
(Fig. \ref{fig:CIB-CMB}). 

With a possible future CIB experiment of (sub)degree angular
resolution, the CIB-SZ cross correlation can be measured. The SZ
effect is sensitive to various thermal heating processes such as SN
feedback and quasar feedback.  The  O and B star
formation  responsible for SCIB is also responsible for SN feedback on
the intergalactic medium,
so the measurement of CIB-SZ cross correlation helps to 
constrain the role of SN feedback on the SZ effect. 

\subsection{Observational feasibility}
Since we will utilize  (almost) full sky data, we expect a factor of
3-4 decrease in the 
sample variance with respect to current WMAP+galaxy
measurement. Furthermore, Interplanetary dust does not correlate with
CMB foregrounds. The main obstacle of CIB-CMB cross correlation
measurement comes from the  
correlations of CIB and CMB galactic foregrounds, mainly the galactic
dust foregrounds. But such correlations concentrate on large
scales. The power spectrum of CIB galactic dust foreground scales as
$C_l\propto l^{-3}$. CMB 
galactic  foregrounds have a combined power spectrum $C_l\propto  l^{-2}$
\citep{Bennett03}. These foreground correlations degrade the detection
of the ISW  effect, but has only minor effect on the detection of the
SZ effect. Furthermore, one can  avoid the galactic plane to minimize
foregrounds.

We can estimate the upper limits of the systematic errors caused by
foreground correlations. For the WMAP W band, the only non-negligible 
foreground is galactic dust emission  with a flat power spectrum $C_l
l^2/(2\pi)\simeq 50\mu\rm k^2$ (Fig. 10, \citet{Bennett03}). This
foreground strongly correlates with the
CIB galactic foreground, which has a power spectrum $C_l\simeq ( 10\ {\rm nW
m^{-2} sr^{-1}})^2 l^{-3}$ (\S \ref{sec:error}). Without any
foreground removal, the systematic error introduced by foreground
correlations 
prohibits the CIB-CMB correlation measurement at $l\la 20$, but it
still leaves a window at $20\la l\la 60$ for the CIB-ISW cross
correlation detection. Its effect to the CIB-SZ cross correlation
measurement at $l\ga 200$ is negligible  (Fig. \ref{fig:CIB-CMB}).

Current foreground removal method using multi-frequency maps can
reduce CMB foreground contaminations in the CMB power spectrum by more
than $90\%$. For cleaned  WMAP map at W band, foreground only
contributes $0.4\%$ of the CMB power spectrum
\citep{Bennett03}. Since maps used for foreground removal correlate
with SCIB, in principle, the foreground removal process may introduce false
correlation signal with SCIB or lose correlation signal. But The ISW
contribution to CMB at $l\sim 10$ and the SZ contribution to CMB at
$l\ga 300$ are much larger than $0.4\%$, thus the CMB signal responsible
for CMB-CIB correlation do survive after the foreground removal
process. so we can neglect such complexity in the cleaned map. Thus
using cleaned CMB map, one expect a factor of $\sim 10$ reduction in the
systematic errors. 

FIRAS has an angular resolution $\sim 5^{\circ}$ and thus can probe
$l\sim 20$. Around such scales, the statistical error of 
CIB-CMB correlation measurement is $\simeq \sqrt{2/(2l+1)\Delta l}\la 10\%$ and the
systematic error is $\la 10\%$ using cleaned WMAP W band map.  So,
combining WMAP and FIRAS, one
expects $\sim 20\%$ accuracy in the CMB-CIB cross correlation measurement.

\section{Conclusion}
Cosmic infrared background (CIB) traces the large scale structure of the
universe and contains plenty of information of the star formation history. But
overwhelming foregrounds prohibit its precision measurement.
By cross correlating cosmic infrared background with galaxies, one can 
eliminate CIB foregrounds, minimize and localize background contaminations and
directly obtain redshift information of CIB sources from galaxy
photometric redshift. Since one does not 
rely on spectral information to extract redshift information of CIB sources,
one only needs 
to estimate the integrated CIB, which is directly determined by star formation
rate through the energy conservation and is
thus fairly model independent.

The CIB-galaxy cross correlation at degree scale is $\sim 10\%$ and can be
measured with $10\%$ accuracy.  We estimated that, the cross correlation could
enable a direct and 
statistically robust measurement of star formation rate with $10\%$ to
$z\lesssim 1.5$.  With our model based on the integrated CIB intensity,  we
predict that the cross correlation between the 
CIB and the CMB is about $0.5 \mu$K at $10^{\circ}$ scale (the ISW effect), changes sign at several
degree scale and reaches $\sim -0.3 \mu$K at one degree scale (the SZ effect). We
argue that  this  cross correlation is 
observationally feasible.  Such measurement would
constrain the  amount  
of dark energy by the ISW effect, the amount of the thermal energy of the
universe by the 
SZ effect and provide further constraint on the clustering property of SFG and
make our CIB model self consistent. 

{\it Acknowledgments}: The author thanks an anonymous referee for many
helpful comments. The author thanks Albert Stebbins and Lam Hui for
helpful discussion and thanks Dale J. Fixsen for explaining
FIRAS beam. This work was supported by the DOE and the NASA grant
NAG 5-10842 at Fermilab.

\end{document}